\documentclass[twocolumn,letterpaper,aps,prl,superscriptaddress,showpacs,floatfix]{revtex4}

\usepackage{graphicx}	

\usepackage{xspace}	

\usepackage{amsmath}

\newcommand{\pt}{\mbox{$p_T$}\xspace}

\newcommand{\Ncoll}{\mbox{$N_{\rm coll}$}\xspace}

\newcommand{\sqsn}{\mbox{$\sqrt{s_{_{NN}}}$}}

\newcommand{\rdau}{\mbox{$R_{d{\rm Au}}$}\xspace}

\newcommand{\pp}{\mbox{$p$+$p$}\xspace}

\newcommand{\dau}{\mbox{$d+$Au}\xspace}  

\newcommand{\midn}{\mbox{$\left|\eta\right|<0.35$}\xspace}

\newcommand{\jpsi}{\mbox{$J/\psi$}\xspace}  
\newcommand{\psip}{\mbox{$\psi'$}\xspace} 
\newcommand{\chic}{\mbox{$\chi_c$}\xspace}
\newcommand{\ccbar}{\mbox{$c\overline{c}$}\xspace}
\newcommand{\jpsiee}{\mbox{$J/\psi\rightarrow e^+e^-$}\xspace}  
\newcommand{\psipee}{\mbox{$\psi'\rightarrow e^+e^-$}\xspace} 
\newcommand{\fdchic}[1]{\mbox{$F_{\chi_c\rightarrow J/\psi}^{#1}$}\xspace}
\newcommand{\chicjpsig}{\mbox{$\chi_c\rightarrow J/\psi + \gamma$}\xspace}

\begin{document}

\title{ Nuclear Modification of $\psi'$, $\chi_c$ and $J/\psi$ Production 
\\ in $d$+Au Collisions at $\sqrt{s_{_{NN}}}=\;200$~GeV }

\newcommand{\abilene}{Abilene Christian University, Abilene, Texas 79699, USA}
\newcommand{\augie}{Department of Physics, Augustana College, Sioux Falls, South Dakota 57197, USA}
\newcommand{\banaras}{Department of Physics, Banaras Hindu University, Varanasi 221005, India}
\newcommand{\barc}{Bhabha Atomic Research Centre, Bombay 400 085, India}
\newcommand{\baruch}{Baruch College, City University of New York, New York, New York, 10010 USA}
\newcommand{\bnlcoll}{Collider-Accelerator Department, Brookhaven National Laboratory, Upton, New York 11973-5000, USA}
\newcommand{\bnlphys}{Physics Department, Brookhaven National Laboratory, Upton, New York 11973-5000, USA}
\newcommand{\caucr}{University of California - Riverside, Riverside, California 92521, USA}
\newcommand{\charlesczech}{Charles University, Ovocn\'{y} trh 5, Praha 1, 116 36, Prague, Czech Republic}
\newcommand{\chonbuk}{Chonbuk National University, Jeonju, 561-756, Korea}
\newcommand{\ciae}{Science and Technology on Nuclear Data Laboratory, China Institute of Atomic Energy, Beijing 102413, P.~R.~China}
\newcommand{\cns}{Center for Nuclear Study, Graduate School of Science, University of Tokyo, 7-3-1 Hongo, Bunkyo, Tokyo 113-0033, Japan}
\newcommand{\colorado}{University of Colorado, Boulder, Colorado 80309, USA}
\newcommand{\columbia}{Columbia University, New York, New York 10027 and Nevis Laboratories, Irvington, New York 10533, USA}
\newcommand{\czechtech}{Czech Technical University, Zikova 4, 166 36 Prague 6, Czech Republic}
\newcommand{\dapnia}{Dapnia, CEA Saclay, F-91191, Gif-sur-Yvette, France}
\newcommand{\elte}{ELTE, E{\"o}tv{\"o}s Lor{\'a}nd University, H - 1117 Budapest, P{\'a}zm{\'a}ny P. s. 1/A, Hungary}
\newcommand{\ewha}{Ewha Womans University, Seoul 120-750, Korea}
\newcommand{\fit}{Florida Institute of Technology, Melbourne, Florida 32901, USA}
\newcommand{\fsu}{Florida State University, Tallahassee, Florida 32306, USA}
\newcommand{\gsu}{Georgia State University, Atlanta, Georgia 30303, USA}
\newcommand{\hiroshima}{Hiroshima University, Kagamiyama, Higashi-Hiroshima 739-8526, Japan}
\newcommand{\ihepprot}{IHEP Protvino, State Research Center of Russian Federation, Institute for High Energy Physics, Protvino, 142281, Russia}
\newcommand{\illuiuc}{University of Illinois at Urbana-Champaign, Urbana, Illinois 61801, USA}
\newcommand{\inrras}{Institute for Nuclear Research of the Russian Academy of Sciences, prospekt 60-letiya Oktyabrya 7a, Moscow 117312, Russia}
\newcommand{\instpasczech}{Institute of Physics, Academy of Sciences of the Czech Republic, Na Slovance 2, 182 21 Prague 8, Czech Republic}
\newcommand{\isu}{Iowa State University, Ames, Iowa 50011, USA}
\newcommand{\jaea}{Advanced Science Research Center, Japan Atomic Energy Agency, 2-4 Shirakata Shirane, Tokai-mura, Naka-gun, Ibaraki-ken 319-1195, Japan}
\newcommand{\jyvaskyla}{Helsinki Institute of Physics and University of Jyv{\"a}skyl{\"a}, P.O.Box 35, FI-40014 Jyv{\"a}skyl{\"a}, Finland}
\newcommand{\kek}{KEK, High Energy Accelerator Research Organization, Tsukuba, Ibaraki 305-0801, Japan}
\newcommand{\korea}{Korea University, Seoul, 136-701, Korea}
\newcommand{\kurchatov}{Russian Research Center ``Kurchatov Institute", Moscow, 123098 Russia}
\newcommand{\kyoto}{Kyoto University, Kyoto 606-8502, Japan}
\newcommand{\labllr}{Laboratoire Leprince-Ringuet, Ecole Polytechnique, CNRS-IN2P3, Route de Saclay, F-91128, Palaiseau, France}
\newcommand{\lahorelums}{Physics Department, Lahore University of Management Sciences, Lahore, Pakistan}
\newcommand{\lawllnl}{Lawrence Livermore National Laboratory, Livermore, California 94550, USA}
\newcommand{\losalamos}{Los Alamos National Laboratory, Los Alamos, New Mexico 87545, USA}
\newcommand{\lpc}{LPC, Universit{\'e} Blaise Pascal, CNRS-IN2P3, Clermont-Fd, 63177 Aubiere Cedex, France}
\newcommand{\lund}{Department of Physics, Lund University, Box 118, SE-221 00 Lund, Sweden}
\newcommand{\maryland}{University of Maryland, College Park, Maryland 20742, USA}
\newcommand{\mass}{Department of Physics, University of Massachusetts, Amherst, Massachusetts 01003-9337, USA }
\newcommand{\michigan}{Department of Physics, University of Michigan, Ann Arbor, Michigan 48109-1040, USA}
\newcommand{\muenster}{Institut fur Kernphysik, University of Muenster, D-48149 Muenster, Germany}
\newcommand{\muhlenberg}{Muhlenberg College, Allentown, Pennsylvania 18104-5586, USA}
\newcommand{\myongji}{Myongji University, Yongin, Kyonggido 449-728, Korea}
\newcommand{\nagasaki}{Nagasaki Institute of Applied Science, Nagasaki-shi, Nagasaki 851-0193, Japan}
\newcommand{\newmex}{University of New Mexico, Albuquerque, New Mexico 87131, USA }
\newcommand{\nmsu}{New Mexico State University, Las Cruces, New Mexico 88003, USA}
\newcommand{\ohio}{Department of Physics and Astronomy, Ohio University, Athens, Ohio 45701, USA}
\newcommand{\ornl}{Oak Ridge National Laboratory, Oak Ridge, Tennessee 37831, USA}
\newcommand{\orsay}{IPN-Orsay, Universite Paris Sud, CNRS-IN2P3, BP1, F-91406, Orsay, France}
\newcommand{\peking}{Peking University, Beijing 100871, P.~R.~China}
\newcommand{\pnpi}{PNPI, Petersburg Nuclear Physics Institute, Gatchina, Leningrad region, 188300, Russia}
\newcommand{\riken}{RIKEN Nishina Center for Accelerator-Based Science, Wako, Saitama 351-0198, Japan}
\newcommand{\rikjrbrc}{RIKEN BNL Research Center, Brookhaven National Laboratory, Upton, New York 11973-5000, USA}
\newcommand{\rikkyo}{Physics Department, Rikkyo University, 3-34-1 Nishi-Ikebukuro, Toshima, Tokyo 171-8501, Japan}
\newcommand{\saopaulo}{Universidade de S{\~a}o Paulo, Instituto de F\'{\i}sica, Caixa Postal 66318, S{\~a}o Paulo CEP05315-970, Brazil}
\newcommand{\stonybrkc}{Chemistry Department, Stony Brook University, SUNY, Stony Brook, New York 11794-3400, USA}
\newcommand{\stonycrkp}{Department of Physics and Astronomy, Stony Brook University, SUNY, Stony Brook, New York 11794-3400, USA}
\newcommand{\tenn}{University of Tennessee, Knoxville, Tennessee 37996, USA}
\newcommand{\titech}{Department of Physics, Tokyo Institute of Technology, Oh-okayama, Meguro, Tokyo 152-8551, Japan}
\newcommand{\tsukuba}{Institute of Physics, University of Tsukuba, Tsukuba, Ibaraki 305, Japan}
\newcommand{\vandy}{Vanderbilt University, Nashville, Tennessee 37235, USA}
\newcommand{\waseda}{Waseda University, Advanced Research Institute for Science and Engineering, 17 Kikui-cho, Shinjuku-ku, Tokyo 162-0044, Japan}
\newcommand{\weizmann}{Weizmann Institute, Rehovot 76100, Israel}
\newcommand{\wigner}{Institute for Particle and Nuclear Physics, Wigner Research Centre for Physics, Hungarian Academy of Sciences (Wigner RCP, RMKI) H-1525 Budapest 114, POBox 49, Budapest, Hungary}
\newcommand{\yonsei}{Yonsei University, IPAP, Seoul 120-749, Korea}
\affiliation{\abilene}
\affiliation{\augie}
\affiliation{\banaras}
\affiliation{\barc}
\affiliation{\baruch}
\affiliation{\bnlcoll}
\affiliation{\bnlphys}
\affiliation{\caucr}
\affiliation{\charlesczech}
\affiliation{\chonbuk}
\affiliation{\ciae}
\affiliation{\cns}
\affiliation{\colorado}
\affiliation{\columbia}
\affiliation{\czechtech}
\affiliation{\dapnia}
\affiliation{\elte}
\affiliation{\ewha}
\affiliation{\fit}
\affiliation{\fsu}
\affiliation{\gsu}
\affiliation{\hiroshima}
\affiliation{\ihepprot}
\affiliation{\illuiuc}
\affiliation{\inrras}
\affiliation{\instpasczech}
\affiliation{\isu}
\affiliation{\jaea}
\affiliation{\jyvaskyla}
\affiliation{\kek}
\affiliation{\korea}
\affiliation{\kurchatov}
\affiliation{\kyoto}
\affiliation{\labllr}
\affiliation{\lahorelums}
\affiliation{\lawllnl}
\affiliation{\losalamos}
\affiliation{\lpc}
\affiliation{\lund}
\affiliation{\maryland}
\affiliation{\mass}
\affiliation{\michigan}
\affiliation{\muenster}
\affiliation{\muhlenberg}
\affiliation{\myongji}
\affiliation{\nagasaki}
\affiliation{\newmex}
\affiliation{\nmsu}
\affiliation{\ohio}
\affiliation{\ornl}
\affiliation{\orsay}
\affiliation{\peking}
\affiliation{\pnpi}
\affiliation{\riken}
\affiliation{\rikjrbrc}
\affiliation{\rikkyo}
\affiliation{\saopaulo}
\affiliation{\stonybrkc}
\affiliation{\stonycrkp}
\affiliation{\tenn}
\affiliation{\titech}
\affiliation{\tsukuba}
\affiliation{\vandy}
\affiliation{\waseda}
\affiliation{\weizmann}
\affiliation{\wigner}
\affiliation{\yonsei}
\author{A.~Adare} \affiliation{\colorado}
\author{C.~Aidala} \affiliation{\mass} \affiliation{\michigan}
\author{N.N.~Ajitanand} \affiliation{\stonybrkc}
\author{Y.~Akiba} \affiliation{\riken} \affiliation{\rikjrbrc}
\author{H.~Al-Bataineh} \affiliation{\nmsu}
\author{J.~Alexander} \affiliation{\stonybrkc}
\author{A.~Angerami} \affiliation{\columbia}
\author{K.~Aoki} \affiliation{\kyoto} \affiliation{\riken}
\author{N.~Apadula} \affiliation{\stonycrkp}
\author{Y.~Aramaki} \affiliation{\cns} \affiliation{\riken}
% ...
\collaboration{PHENIX Collaboration} \noaffiliation

\date{\today}

\begin{abstract}

We present results for three charmonia states ($\psi'$, $\chi_c$, and 
$J/\psi$) in $d$+Au collisions at $|y|<0.35$ and 
$\sqrt{s_{_{NN}}}=200$~GeV.  We find that the modification of the $\psi'$ 
yield relative to that of the $J/\psi$ scales approximately with charged 
particle multiplicity at midrapidity across $p+$A, $d$+Au, and $A$$+$$A$ 
results from the Super Proton Synchrotron and the Relativistic Heavy Ion 
Collider.  In large-impact-parameter collisions we observe a similar 
suppression for the $\psi'$ and $J/\psi$, while in small-impact-parameter 
collisions the more weakly bound $\psi'$ is more strongly suppressed.  
Owing to the short time spent traversing the Au nucleus, the larger 
$\psi'$ suppression in central events is not explained by an increase of 
the nuclear absorption due to meson formation time effects.

\end{abstract}

\pacs{25.75.Dw} 
	

\maketitle


Understanding the evolution of heavy quark-antiquark pairs into bound 
color singlet quarkonium states represents a challenge within QCD.  An 
excellent tool for probing the time scale for this evolution is the 
measurement of production rates for multiple quarkonium states, with 
different physical sizes and binding energies, in proton (or deuteron) - 
nucleus collisions.  The evolving quark-antiquark pair must traverse the 
target nucleus, and by varying the path length in the nucleus one can 
probe this time scale.

Measurements of \jpsi and \psip production rates at 
$\sqrt{s_{_{NN}}} = 38.7$~GeV, as a function of Feynman-$x$ 
($x_F$), in proton-nucleus collisions by E866/NuSea~\cite{Leitch:1999ea} 
show a greater suppression of \psip production compared to \jpsi 
production near $x_F\approx$~0, and a comparable suppression for $x_F>0$. 
Similar measurements by NA50~\cite{Alessandro:2006jt} at 
\mbox{\sqsn~=~27.4 GeV} and $x_F\approx$~0 show a stronger suppression of 
\psip production, compared to \jpsi production, for larger nuclei. This 
has been interpreted as an effect of the charmonia formation 
time~\cite{Arleo:1999af}. When the time spent traversing the nucleus by 
the \ccbar pair becomes longer than the charmonia formation time, the 
larger \psip meson will be further suppressed by a larger nuclear breakup 
effect. It is critical to test these assumptions at the collision energies 
provided by the Relativistic Heavy Ion Collider (RHIC), where the time 
spent traversing the nucleus is expected to be much shorter than this 
formation time. Also, the binding energy of the the \psip 
($\approx$~0.05~GeV) is significantly smaller than that of the \chic 
($\approx$~0.20~GeV) or \jpsi 
($\approx$~0.64~GeV)~\cite{Satz:2005hx}, and 
may play an important role in understanding the effects of producing 
quarkonia in a nuclear target.

The PHENIX experiment has previously reported measurements of \jpsi 
production rates in \dau collisions at \sqsn~=~200 GeV using data 
collected in 2008~\cite{Adare:2010fn, Adare:2012qf}. Here we present 
measurements of \psip production rates, as well as the fraction of \jpsi 
yield which comes from \chic decays, in \dau collisions at midrapidity 
from the same data set. Using the corresponding measurements in \pp 
collisions by PHENIX~\cite{Adare:2011vq}, we construct the nuclear 
modification factor, \rdau, for \psip and \chic production and compare it 
with the measurements of the \jpsi \rdau at the same energy.

The PHENIX detector is described in detail in Ref.~\cite{Adcox:2003zm}. 
The data presented here were collected using the two PHENIX central arms, 
each of which detect electrons, photons, and hadrons over \midn and 
$\Delta\phi=\frac{\pi}{2}$. The \dau data used in this analysis were 
recorded using a minimum bias (MB) trigger in coincidence with an 
additional electron Level-1 trigger. The MB trigger requires at least one 
hit in each of the two beam-beam counters (BBCs) covering $3<|\eta|<3.9$. 
This MB selection covers 88$\pm$4\% of the total \dau inelastic cross 
section of 2.26 barns~\cite{White:2005kp}. The electron trigger requires a 
minimum energy deposited in any group of 2$\times$2 towers in the 
Electromagnetic Calorimeter and an associated hit in the Ring 
Imaging \v{C}erenkov counter. Thresholds of 600 and 800 MeV were 
used, each for roughly half of the data sample. The data set represents 
analyzed integrated luminosities of 62.7 and 66.2 nb$^{-1}$ for the \psip 
and \chic analyses respectively.

The \psip invariant yield is calculated as
\begin{equation}
B_{ee}\frac{dN_{\psi'}}{dy}=\frac{cN_{\psi'}}{N_{\rm MB}\epsilon A\Delta y},
\label{eq:inv_yield}
\end{equation} 
where $B_{ee}$ is the $\psip\rightarrow e^+e^-$ branching ratio, 
$N_{\psi'}$ is the measured \psipee yield, $N_{\rm MB}$ is the number of 
sampled MB events, and $\Delta y$ is the width of the rapidity bin. 
A {\sc geant}-3 based model of the PHENIX detector combined with 
measurements of the momentum dependence of the single electron trigger 
efficiency, as described in Ref.~\cite{Adare:2012qf}, is used to calculate 
the product of the acceptance and efficiency, $\epsilon A$, which includes 
the Level-1 trigger efficiency. This model is also used to estimate the 
detector effects on the simulated signal and background line shapes when 
fitting the measured dielectron signal. Following the procedures described 
in Ref.~\cite{Adare:2012qf}, $\epsilon A$ is found to have an average 
value of 0.91\% with a relative systematic uncertainty of 6.4\%. The 
correction factor $c$ accounts for the trigger and centrality bias present 
in events which contain a hard scattering~\cite{Adare:2012qf}. The track 
multiplicity dependence of the reconstruction efficiency is negligible in 
\dau collisions, and a 1\% systematic uncertainty was assigned based on 
the \jpsi studies performed in Ref.~\cite{Adare:2012qf}.

The \psipee yield is extracted from fits to the unlike-sign ($e^+e^-$) 
invariant mass distribution, after the subtraction of the like-sign 
($e^+e^++e^-e^-$) background, where at least one of the electrons fired 
the Level-1 trigger. The fit is performed over the mass range 
$2.0<M_{ee}\,[{\rm GeV}/c^2]<5.5$, and includes line shapes for \jpsiee 
and \psipee decays, as well as the remaining correlated background from 
open heavy flavor and Drell-Yan decays.

The \jpsi and \psip line shapes include the natural line shape, smeared 
based on the PHENIX mass resolution, and radiative decays 
($\jpsi\rightarrow e^+e^-\gamma$ for $E_{\gamma}>100$ MeV), using 
calculations of the mass distribution from QED~\cite{Spiridonov:2004mp}. 
The line shape for Drell-Yan decays was generated using 
{\sc pythia}-6~\cite{Sjostrand:2006za}. Line shapes for open heavy flavor 
decays were generated using three different MC generators, including 
{\sc pythia}-6 in both hard scattering and forced charm (or bottom) 
production modes as well as the MC@NLO generator~\cite{Frixione:2003ei}.

\begin{figure}[ht]
\includegraphics[width=1.0\linewidth]{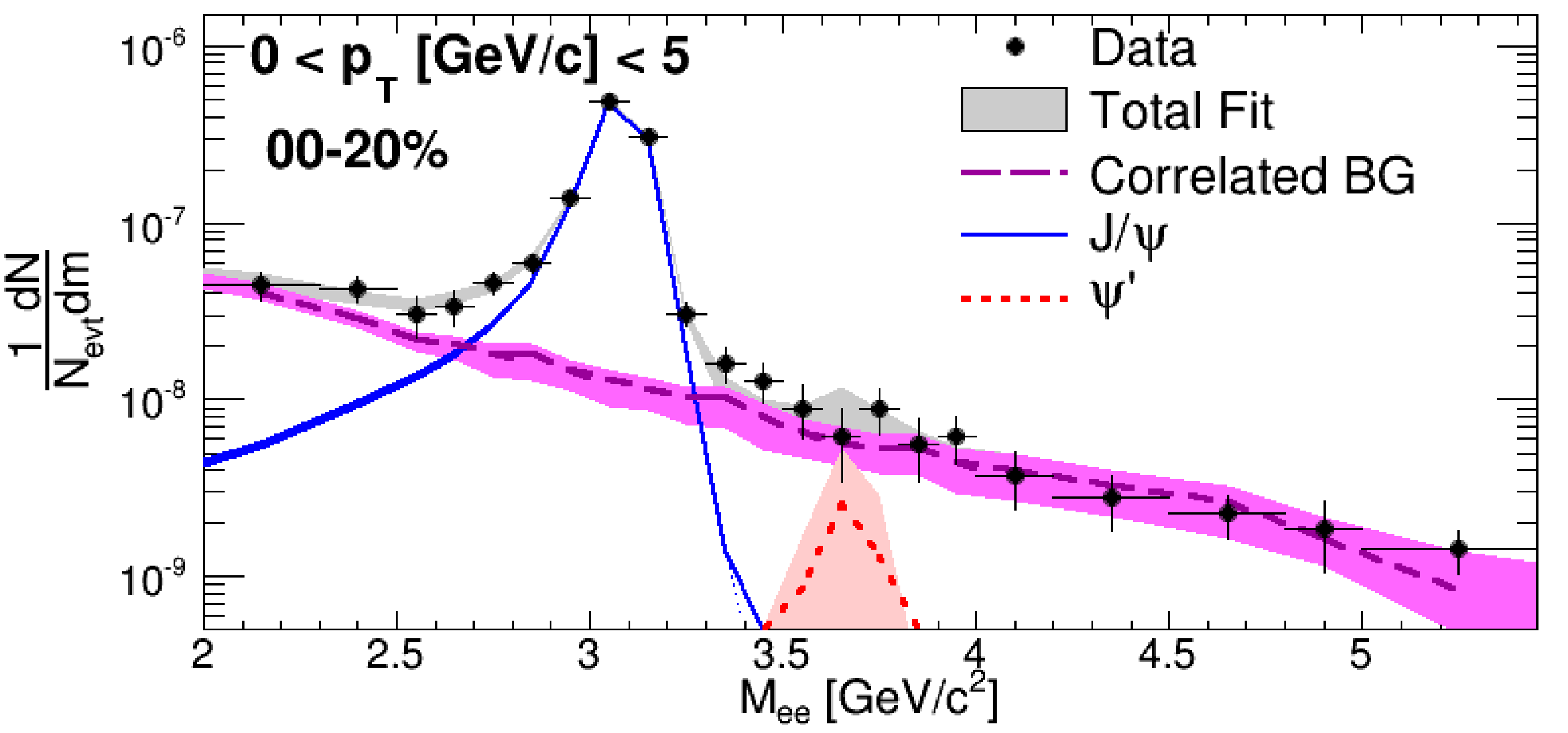} 
\includegraphics[width=1.0\linewidth]{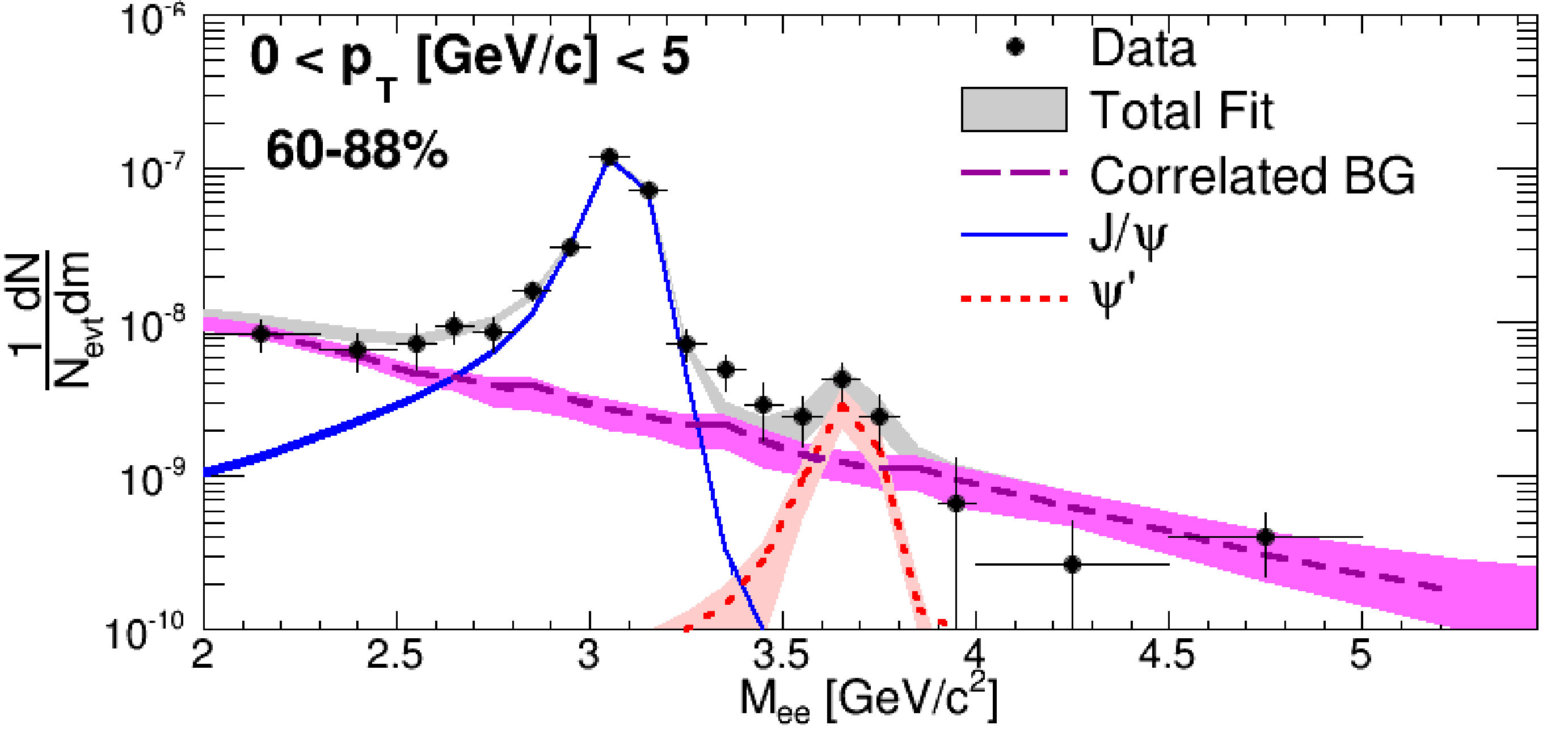} 
\caption{\label{fig:psip_massdist}(Color online) 
The $e^+e^-$ mass distribution, after like-sign subtraction, for 0--20\% 
(Top) and 60--88\% (Bottom) \dau collisions. The line shapes are those fit 
to the data in order to extract the \psip yield.
}
\end{figure}

After applying the detector acceptance and efficiency effects, the line 
shapes are fit to the invariant mass distributions. It was found that the 
heavy flavor line shapes generated using {\sc pythia}-6 set to hard 
scattering mode gave the lowest $\chi^2$ per degree of freedom (68.5/68), 
while those generated using {\sc pythia}-6 set to charm(bottom) production 
as well as those generated using MC@NLO provided slightly poorer 
agreements with a $\chi^2$ per degree of freedom of 79.1/68 and 83.4/68 
respectively. The different line shapes resulted in changes in the 
extracted \psip yield of less than 20\% in peripheral events. In central 
events there is a very small signal which varies by up to 83\% within the 
different assumed shapes. In all cases, the continuum line shapes were 
generated for \pp collisions, and may be modified in \dau collisions. The 
effect of nuclear shadowing on the Drell-Yan and open heavy flavor line 
shapes using the EPS09s parametrization~\cite{Helenius:2012wd} was found 
to change the extracted \psip yield by less than 5\%.

Figure~\ref{fig:psip_massdist} shows the results of the fit for central 
and peripheral \dau collisions. The shaded bands represent the combined 
uncertainty in the fit normalizations, as well as changes in the shape of 
the correlated background obtained using the three different sets of open 
heavy flavor line shapes.

The resulting invariant yields are used, in conjunction with the measured 
values in \pp collisions~\cite{Adare:2011vq}, to calculate the nuclear 
modification factor, \rdau. The \psip \rdau is calculated as
\begin{equation}
R_{dAu}^{\psi'} = \frac{dN^{dAu}_{\psi'}/dy}{\Ncoll dN^{pp}_{\psi'}/dy},
\label{eq:rdau}
\end{equation}
where \Ncoll is the mean number of nucleon-nucleon collisions, and 
$dN^{dAu}_{\psi'}/dy$ and $dN^{pp}_{\psi'}/dy$ are the measured invariant 
yields in \dau and \pp collisions, respectively. The value of \Ncoll is 
calculated using a Glauber Monte Carlo model coupled with a simulation of 
the PHENIX BBC response (see~\cite{Adare:2012qf} for details). We find a 
value for the \psip nuclear modification factor of 
$R_{dAu}^{\psi'}=0.54\pm0.11(\textrm{stat})^{+0.19}_{-0.16}(\textrm{\rm 
syst})$ in 0--100\% centrality integrated \dau collisions.

The feed-down fraction of the inclusive \jpsi yield from \chic decays in 
\dau collisions (\fdchic{dAu}) is measured via the $\chicjpsig\rightarrow 
e^+e^-+\gamma$ decay channel, where the $e^+e^-\gamma$ is fully 
reconstructed in the PHENIX central arms. The procedure for extracting 
\fdchic{dAu} is the same as that presented for \pp collisions 
in~\cite{Adare:2011vq} for a data sample of comparable statistical 
precision. The final feed-down fraction is found to be 
$F_{\chi_c\rightarrow 
J/\psi}^{dAu}=0.32\pm0.09(\textrm{stat})\pm0.03(\textrm{\rm syst})$.

Using the measured feed-down fraction in \pp collisions and the \jpsi 
\rdau, the \chic \rdau is calculated as
\begin{equation}
R^{\chi_c}_{dAu}=R^{J/\psi}_{dAu}\times\frac{\fdchic{dAu}}{\fdchic{pp}}.
\label{eq:chic_rdau}
\end{equation}
The nuclear modification of \chic production in \dau collisions is found 
to be $R_{dAu}^{\chi_c}=0.77\pm0.41(\textrm{stat})\pm0.18(\textrm{\rm 
syst})$.

\begin{figure}[t]
\includegraphics[width=1.0\linewidth]{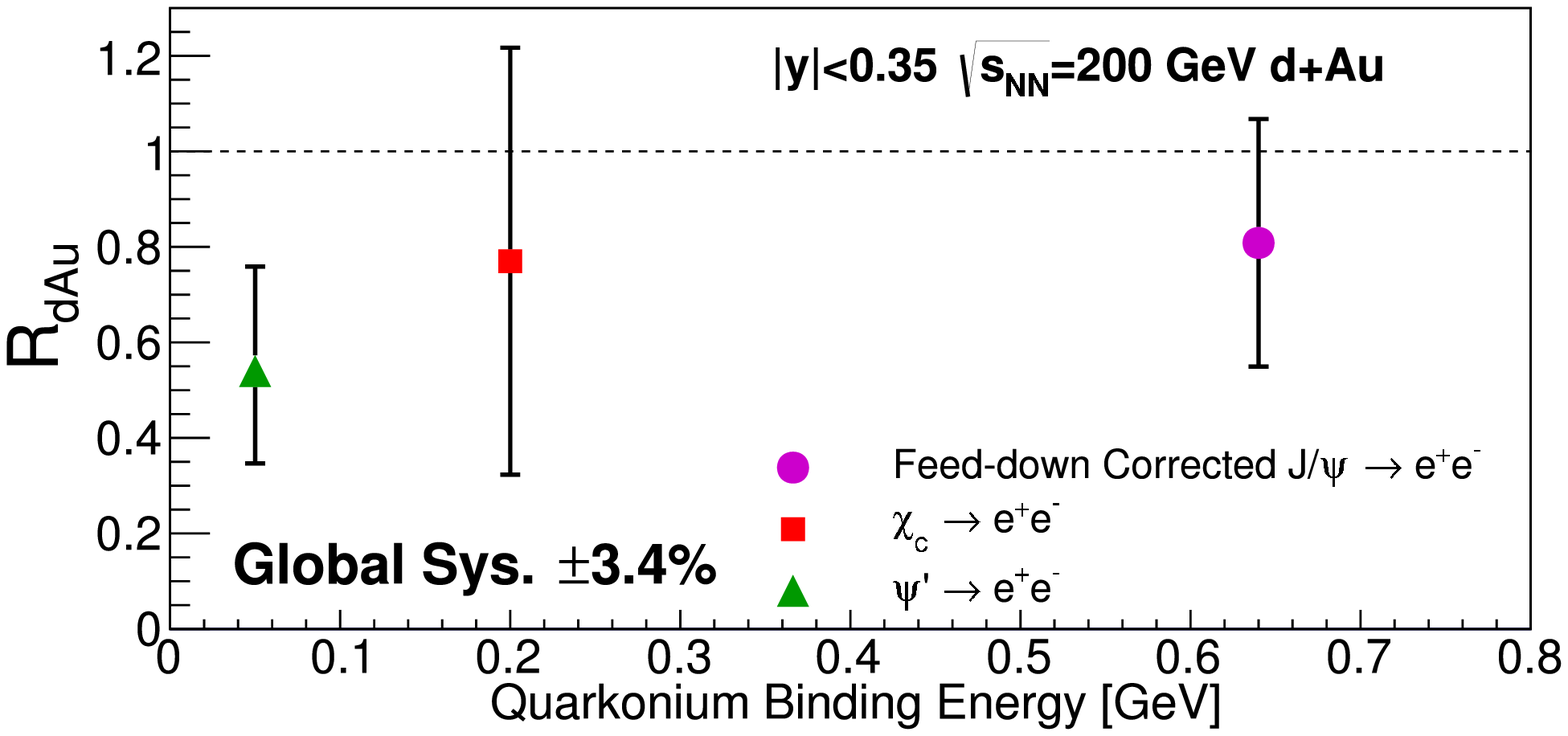}
\caption{\label{fig:charmonia_rdau_BE}(Color online) \
The \psip, \chic, and \jpsi \rdau for 0--100\% centrality integrated \dau 
collisions as a function of the quarkonia binding energy, where the \jpsi 
\rdau has been corrected for the effects of \psip and \chic feed-down. The 
systematic uncertainties which are not correlated between the three points 
and have been added in quadrature with the statistical uncertainties for 
plotting. The common global scale uncertainty is quoted on the plot. The 
binding energies are the differences between the quarkonium masses and the 
open charm threshold, and are taken from Ref.~\cite{Satz:2005hx}.
}
\end{figure}

With the \psip and \chic nuclear modification in hand, it is possible to 
correct the measured modification of inclusive \jpsi production for their 
feed-down effects, thus giving a closer representation of the modification 
of direct \jpsi production. Here we use the \psip and \chic 
feed-down fractions in \pp collisions measured by PHENIX in 
Ref.~\cite{Adare:2011vq}.  The corrected \jpsi modification is calculated 
as 
\begin{equation}
R_{dAu}^{\mathrm{direct}\,J/\psi}=\frac{\left(R_{dAu}^{\mathrm{incl}\,J/\psi}
-F_{\psi'\rightarrow J/\psi}^{pp}R_{dAu}^{\psi'}
-F_{\chi_c\rightarrow J/\psi}^{pp}R_{dAu}^{\chi_c}\right)}
{\left(1-F_{\psi'\rightarrow J/\psi}^{pp}
-F_{\chi_c\rightarrow J/\psi}^{pp}\right)},
\label{eq:jpsi_direct}
\end{equation}
where $R_{dAu}^{\mathrm{incl}\,J/\psi}=0.77\pm0.02(\rm{stat})\pm0.16(\rm{syst})$ 
is the modification of inclusive \jpsi production, reported in 
Ref.~\cite{Adare:2010fn}. This gives a feed-down corrected \jpsi 
modification of 
$R_{dAu}^{\mathrm{direct}\,J/\psi}=0.81\pm0.12(\textrm{stat})\pm0.23(\textrm{\rm 
syst})$. While there still remains a contribution from 
$B\rightarrow\jpsi+X$ decays, its value is expected to be small 
($\approx$~2.7\%~\cite{Cacciari:2005rk}).

\begin{figure}[t]
\includegraphics[width=1.0\linewidth]{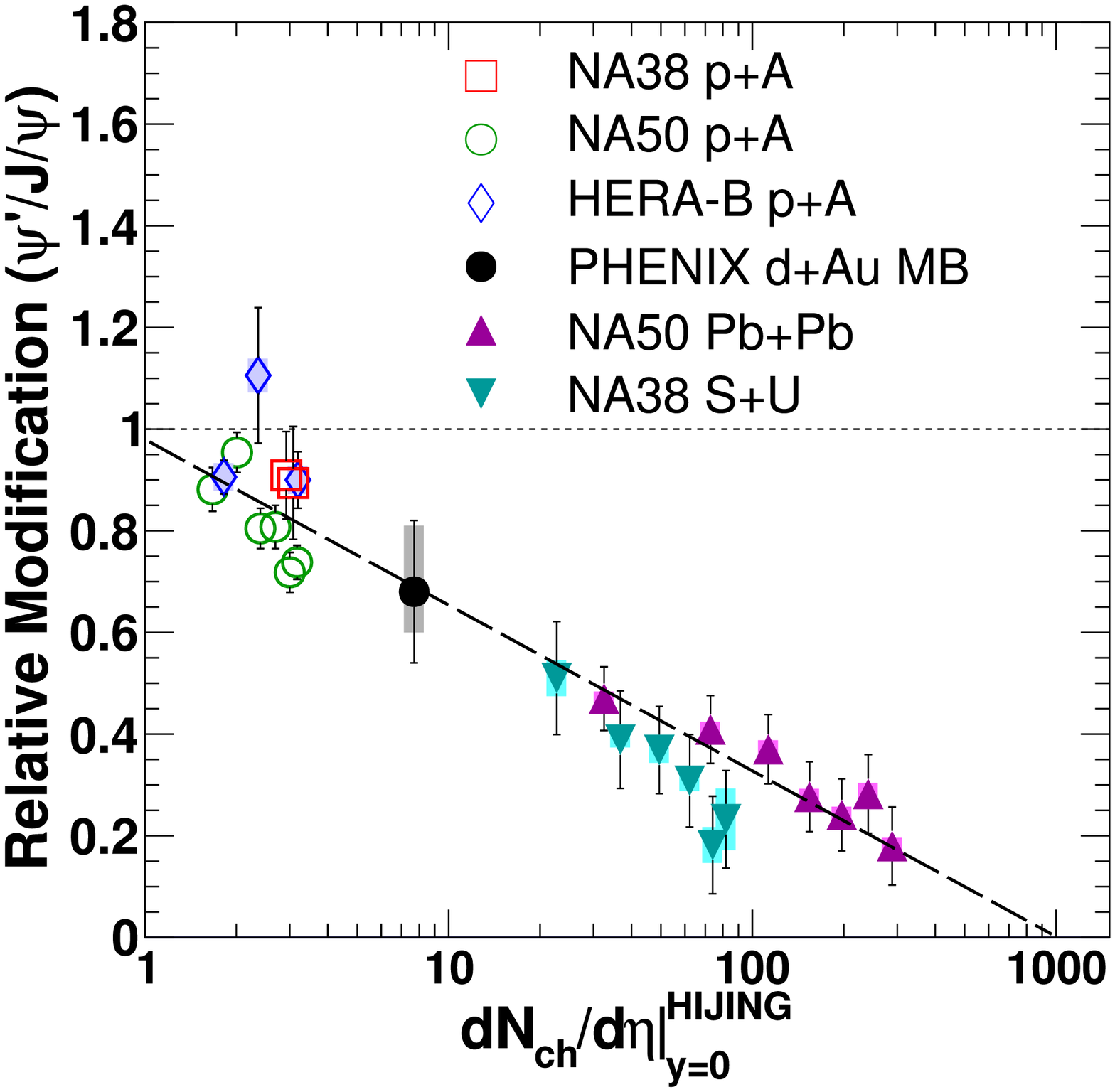}
\caption{\label{fig:psip_relmod_dNchdeta}(Color online) 
The relative modification of the \psip to the \jpsi as a function of 
$dN_{ch}/d\eta|_{y=0}$. The plotted data include (open [red] squares) 
NA38~\cite{Alessandro:2006jt} $p$$+$$A$ at \sqsn~=~19.4 GeV, (open [green] 
circles) NA50~\cite{Abreu:1998wx} $p$$+$$A$ at \sqsn~=~27.4 GeV,(open 
[blue] diamonds) HERA-B~\cite{Abt:2006va} $p$$+$$A$ at \sqsn~=~41.5 with a 
global uncertainty of $\pm$4.4\%, (closed [black] circles) PHENIX \dau at 
\sqsn~=~200 GeV with a global uncertainty of $\pm$24\%, (closed [magenta] 
upward-pointing triangle) NA50~\cite{Alessandro:2006ju} Pb+Pb at 
\sqsn~=~17.2 GeV, and (closed [cyan] downward-pointing triangles) 
NA38~\cite{Alessandro:2006ju} S$+$U at \sqsn~=~19.4 GeV. The SPS and 
HERA-B results are calculated using the extrapolated $p$$+$$p$ \psip to 
\jpsi ratios quoted in the respective references. There is a common global 
uncertainty in the SPS points of 5\% due to the uncertainty in the \pp 
\psip/\jpsi ratio. The dashed line is included only to guide the eye.
}
\end{figure}

Figure~\ref{fig:charmonia_rdau_BE} plots the nuclear modification as a 
function of the quarkonia binding energy. The \chic measurement has large 
statistical and systematic uncertainties, but the \jpsi and \psip 
modifications suggest that there is a decrease in suppression with 
increasing binding energy.

Figure~\ref{fig:psip_relmod_dNchdeta} compares the PHENIX results to data 
taken at different collision energies and species by plotting the relative 
modification of \psip to \jpsi production 
($R_{dAu}^{\psi'}/R_{dAu}^{\mathrm{incl}\,J/\psi}$) as a function of charged 
particle multiplicity. When taking the \psip to \jpsi ratio, a number of 
uncertainties cancel or are reduced, such as the uncertainty in $\epsilon 
A$. Nuclear effects that are common between the \jpsi and \psip (such as 
nuclear shadowing) will also cancel. As there are currently no 
measurements available of $dN_{ch}/d\eta|_{y=0}$ for the majority of the 
data shown in Fig.~\ref{fig:psip_relmod_dNchdeta}, we use 
HIJING~\cite{Gyulassy:1994ew} to calculate the $dN_{ch}/d\eta|_{y=0}$ 
values for all points.  The consistent trend of results at the Super 
Proton Synchrotron (SPS), Hadron-Electron Ring Accelerator (HERA), and 
RHIC, suggests that interactions with final-state hadrons may play a role.
The \psip \rdau is further calculated for different centrality bins 
matched to those used in the previous \jpsi analyses~\cite{Adare:2010fn, 
Adare:2012qf}. \

\begin{figure}[thb]
\includegraphics[width=1.0\linewidth]{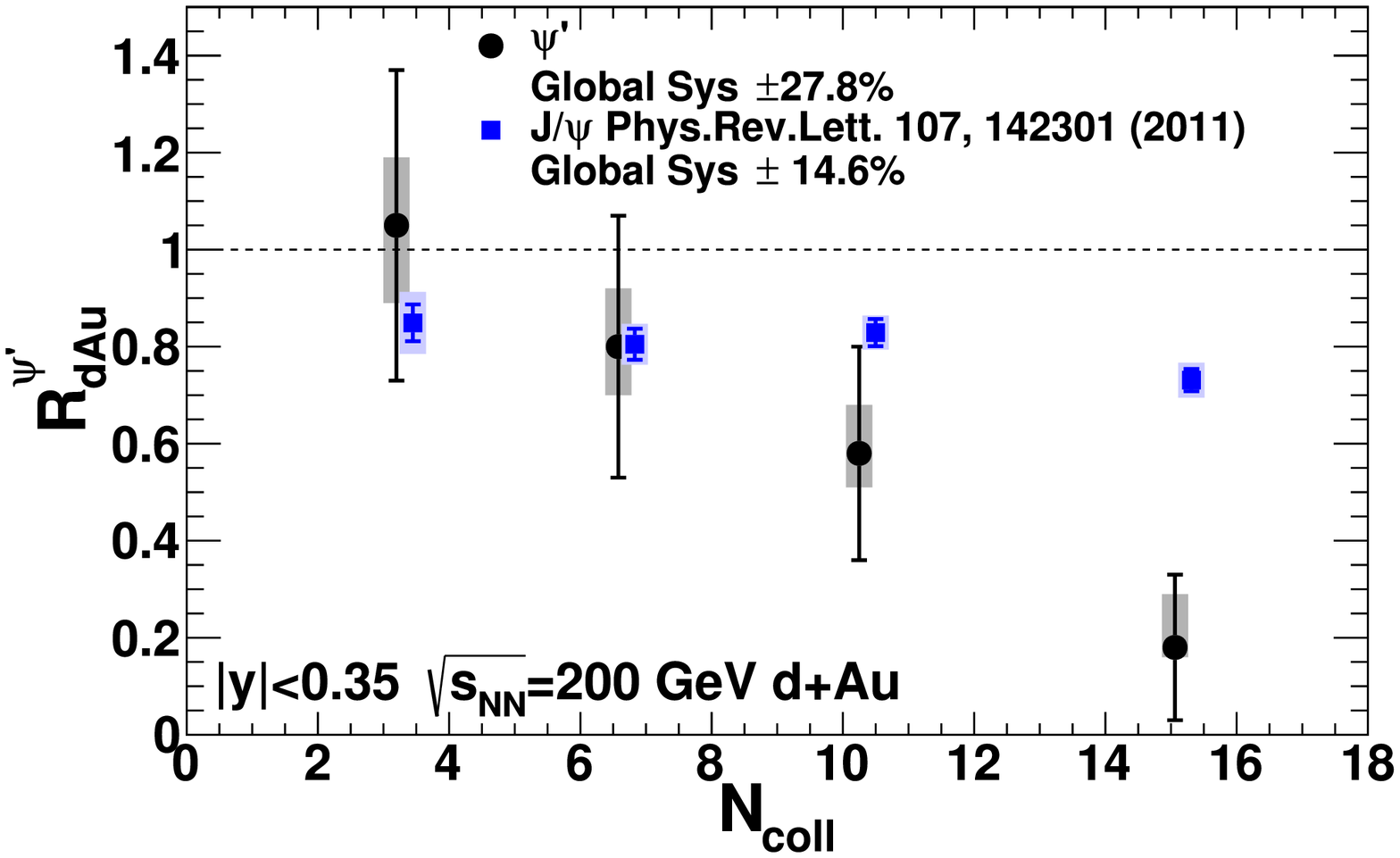}
\caption{\label{fig:psip_rdau_cent}(Color online) 
The \psip nuclear modification factor, \rdau, as a function of \Ncoll. 
Also included are the previously measured \jpsi \rdau as a function of 
\Ncoll~\cite{Adare:2010fn}. Note that the \jpsi \rdau plotted here is not 
corrected for \psip and \chic feed-down, and the \Ncoll values are shifted 
slightly to aid in clarity.
}
\end{figure}

Figure~\ref{fig:psip_rdau_cent} shows \psip \rdau as a function of \Ncoll 
and also shows the previously published \jpsi \rdau~\cite{Adare:2010fn}, 
here integrated over the full rapidity coverage of the central arm.  We 
observe a strong suppression of \psip production with increasing \Ncoll.  
The observed suppression in central \dau collisions (large \Ncoll) is a 
factor of $\approx$~3 times larger than the observed suppression for 
inclusive \jpsi production.

Ref.~\cite{Arleo:1999af} presents a model that explains the lower energy 
E866/NuSea and NA50 results using an expanding color neutral \ccbar pair. 
As the \ccbar expands, it has an increased nuclear absorption due to its 
larger physical size. Once the time spent by the \ccbar pair traversing 
the nucleus becomes larger than the \jpsi formation time, the \psip will 
see a larger nuclear absorption due to its larger size 
($r_0\approx$~0.9~fm 
for the \psip and $r_0\approx$~0.5~fm for the \jpsi~\cite{Satz:2005hx}). 
This explains the transition from a similar level of suppression between 
the \jpsi and \psip at high $x_F$ to a larger suppression of the \psip 
relative to the \jpsi at $x_F\approx$~0 observed by E866/NuSea.

This idea is tested at RHIC energies by calculating the average proper 
time, $\tau$, spent in the nucleus by the quarkonia (or \ccbar precursor). 
This is calculated as $\tau=\beta\langle L \rangle$, where $\langle L 
\rangle$ is the mean thickness of the target nucleus, and $\beta$ is the 
average velocity of the quarkonia in the rest frame of the target nucleus. 
Here the \jpsi \pt is neglected. The $\langle L \rangle$ values for each 
centrality bin are calculated as the average center to edge distance using 
the same Glauber Monte Carlo model used to determine \Ncoll.

\begin{figure}[htb]
\includegraphics[width=1.0\linewidth]{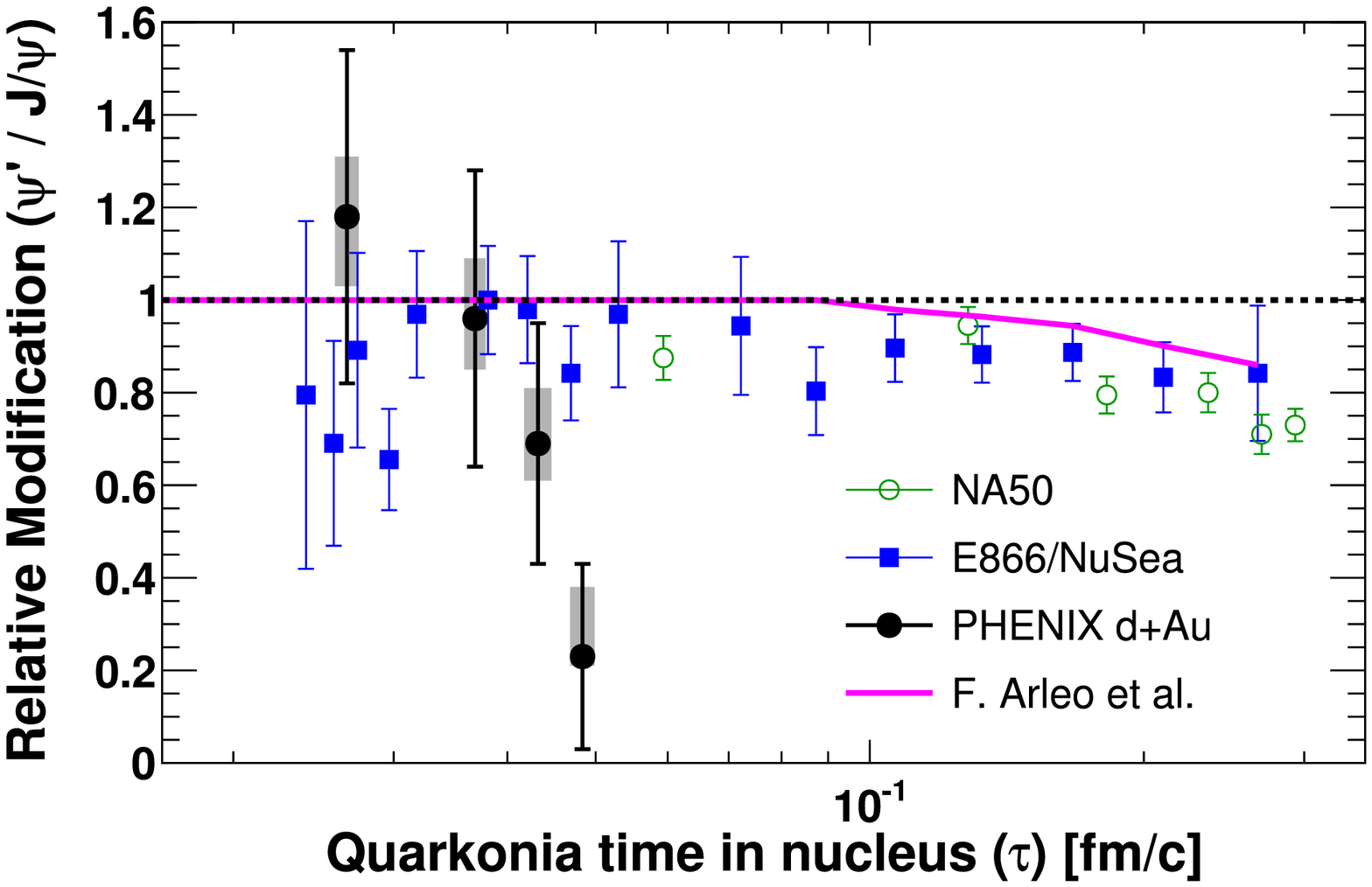}
\caption{\label{fig:psip_relmod_tau}(Color online) 
The relative modification of the \psip to the \jpsi as a function of the 
proper time spent by the quarkonia (or \ccbar precursor) in the nucleus. 
The data include (open [green] circles) NA50~\cite{Abreu:1998wx} $p$$+$$A$ 
at 400 GeV/nn, (closed [blue] squares) E866/NuSea~\cite{Leitch:1999ea} 
$p$$+$$A$ at 800 GeV/nn and (closed [black] circles) PHENIX \dau at 
\sqsn~=~200 GeV which include a global systematic uncertainty of 
$\pm$24\%. The E866/NuSea points are calculated for \psip and \jpsi 
modifications in similar rapidity intervals. The curve is a calculation by 
Arleo {\it et al.}~\cite{Arleo:1999af} discussed in the text.
}
\end{figure}

Figure~\ref{fig:psip_relmod_tau} shows the relative modification of the 
\psip to the \jpsi as a function of $\tau$, where the E866/NuSea and NA50 
results have also been included.  The solid curve is the calculation by 
Arleo {\it et al.}~\cite{Arleo:1999af}, which is consistent with the trends 
observed by E866/NuSea and NA50.

The values of $\tau$ for the PHENIX data are similar to the \ccbar 
formation and color neutralization time of $\approx$~0.05~fm/$c$, and well 
below the \jpsi formation time of 
$\approx$~0.15~fm/$c$~\cite{Arleo:1999af}.  Therefore the model cannot 
explain the strong differential suppression of the \psip in the PHENIX 
data.  We note that Ref.~\cite{McGlinchey:2012bp} shows that the extracted 
break up cross section for the inclusive \jpsi displays a strong departure 
of the E866/NuSea result from $\tau$ scaling below $\approx$~0.05~fm/$c$, 
indicating the presence of different effects that modify charmonium 
production at short time scales. The PHENIX data further indicate that 
there are effects at short crossing time scales that can differentially 
suppress the \psip relative to the \jpsi.

In summary, we have presented measurements of \psip production, as well as 
the \jpsi feed-down fraction from \chic decays, in \dau collisions at 
\sqsn~=~200 GeV. Using the corresponding measurements in \pp collisions, 
we 
have obtained the nuclear modification factor, \rdau, for \psip and \chic 
production. We find that the relative modification of \psip to inclusive 
\jpsi measured by PHENIX follows the same approximate scaling with the 
charged particle multiplicity measured at midrapidity as lower energy 
data. We further find that \psip production is heavily suppressed in 
central \dau collisions relative to \jpsi production. Because the nuclear 
crossing time is very short, this cannot be explained by the difference in 
size of the fully-formed \psip and \jpsi. It instead suggests that there 
is a process occurring on the time scale of \ccbar formation that 
differentially suppresses the \psip.



\begin{acknowledgments}


We thank the staff of the Collider-Accelerator and Physics
Departments at Brookhaven National Laboratory and the staff of
the other PHENIX participating institutions for their vital
contributions.  We acknowledge support from the 
Office of Nuclear Physics in the
Office of Science of the Department of Energy, the
National Science Foundation, Abilene Christian University
Research Council, Research Foundation of SUNY, and Dean of the
College of Arts and Sciences, Vanderbilt University (U.S.A),
Ministry of Education, Culture, Sports, Science, and Technology
and the Japan Society for the Promotion of Science (Japan),
Conselho Nacional de Desenvolvimento Cient\'{\i}fico e
Tecnol{\'o}gico and Funda\c c{\~a}o de Amparo {\`a} Pesquisa do
Estado de S{\~a}o Paulo (Brazil),
Natural Science Foundation of China (P.~R.~China),
Ministry of Education, Youth and Sports (Czech Republic),
Centre National de la Recherche Scientifique, Commissariat
{\`a} l'{\'E}nergie Atomique, and Institut National de Physique
Nucl{\'e}aire et de Physique des Particules (France),
Bundesministerium f\"ur Bildung und Forschung, Deutscher
Akademischer Austausch Dienst, and Alexander von Humboldt Stiftung (Germany),
Hungarian National Science Fund, OTKA (Hungary), 
Department of Atomic Energy and Department of Science and Technology (India), 
Israel Science Foundation (Israel), 
National Research Foundation and WCU program of the 
Ministry Education Science and Technology (Korea),
Ministry of Education and Science, Russian Academy of Sciences,
Federal Agency of Atomic Energy (Russia),
VR and Wallenberg Foundation (Sweden), 
the U.S. Civilian Research and Development Foundation for the
Independent States of the Former Soviet Union, 
the US-Hungarian Fulbright Foundation for Educational Exchange,
and the US-Israel Binational Science Foundation.

\end{acknowledgments}



\end{document}